\documentclass[aps,prb,preprint,floatfix,superscriptaddress]{revtex4-1}

\usepackage{amsmath}
\usepackage{amssymb}
\usepackage{times}
\usepackage{braket}
\usepackage{graphicx}
\usepackage{blindtext}
\usepackage{hyperref}
\usepackage{cleveref}
\usepackage{color}

\renewcommand{\vec}[1]{\mbox{\boldmath$\mathrm{#1}$}}
\def\ind#1{{_{\mathrm{#1}}}}

\newcommand{\dd}{\mathrm{d}}

\begin{document}

\title{Rotating edge-field driven processing of chiral spin textures in racetrack devices}

\author{Alexander F. Sch\"affer}
\email{alexander.schaeffer@physik.uni-halle.de }
\affiliation{Institute of Physics, Martin-Luther-Universit\"at Halle-Wittenberg, D-06120 Halle (Saale), Germany}
\affiliation{Department of Physics, Universit\"at Hamburg, D-20355 Hamburg, Germany}

\author{Pia Siegl}
\author{Martin Stier}
\author{Thore Posske}
\affiliation{I. Institute for Theoretical Physics, Universit\"at Hamburg, D-20355 Hamburg, Germany}

\author{Jamal Berakdar}
\affiliation{Institute of Physics, Martin-Luther-Universit\"at Halle-Wittenberg, D-06120 Halle (Saale), Germany}

\author{Michael Thorwart}
\affiliation{I. Institute for Theoretical Physics, Universit\"at Hamburg, D-20355 Hamburg, Germany}

\author{Roland Wiesendanger}
\affiliation{Department of Physics, Universit\"at Hamburg, D-20355 Hamburg, Germany}

\author{Elena Y. Vedmedenko}
\email[]{vedmeden@physnet.uni-hamburg.de}
\affiliation{Department of Physics, Universit\"at Hamburg, D-20355 Hamburg, Germany}

\date{\today}

\begin{abstract}
	Topologically distinct magnetic structures like skyrmions, domain walls, and the uniformly magnetized state have multiple applications in logic devices, sensors, and as bits of information. One of the most promising concepts for applying these bits is the racetrack architecture controlled by electric currents or magnetic driving fields.	In state-of-the-art racetracks, these fields or currents are applied to the whole circuit.
	Here, we employ micromagnetic and atomistic simulations to establish a concept for racetrack memories free of global driving forces. 
	Surprisingly, we realize that mixed sequences of topologically distinct objects can be created and propagated over far distances exclusively by local rotation of magnetization at the sample boundaries. We reveal the dependence between the chirality of the rotation and the direction of propagation and define the phase space where the proposed procedure can be realized.
	The advantages of this approach are the exclusion of high current and field densities as well as its compatibility with an energy-efficient three-dimensional design. 
\end{abstract}

\maketitle

Magnetic logic devices based on magnetic domain walls (DW) were introduced in 2005\cite{allwood2005DWlogic}, where DWs were driven by rotating magnetic fields in magnetic stripes. Subsequently, the control of the domain walls by electric currents was proposed in memory devices\cite{parkin2008magnetic}. Since that time, both driving mechanisms were refined and extended to chiral objects\cite{sampaio2013nucleation,fernandez2017threeD,luo2019chirally,luo2020current}. 
Recently, we have shown theoretically\cite{vedmedenko2014topologically} that certain topological magnetic structures can be created without the help of global fields or currents, only by imposing time-dependent boundary conditions. Particularly, stable spirals of classical spins of different winding numbers can be created in antiferromagnetic chains by a local rotation of magnetization at the chain ends. The concept has also been applied to chains of quantum spins recently\cite{posske2019winding}.
This one-dimensional model can be applied to multilayered pillars\cite{Cowburn2013Ratchet} or atomistic magnetic chains on substrates\cite{menzel2012chains}.  An appealing idea is to use such constraints to create metastable topological quasiparticles like domain walls (DWs), solitons, or skyrmions, in two- or three-dimensional racetracks. 
A significant decrease in energy consumption could be achieved by using similar time-dependent boundary conditions for further transfer and storage of magnetic objects created alongside the racetrack.
However, it was not clear up to now whether the one-dimensional concept\cite{vedmedenko2014topologically} can be applied to racetracks and whether these magnetic objects - once created - can then be transported and deleted by locally manipulating the boundaries without global currents or fields.
Here, we develop theoretical concepts for the processing of topological magnetic objects in racetracks, including their creation, deletion, and transportation using time-dependent boundary conditions only. We especially consider systems with a relevant Dzyaloshinskii-Moriya interaction (DMI), potentially hosting both chiral DWs and skyrmions.

\section*{Results}
\begin{figure}
	\centering
	\includegraphics[width=\linewidth]{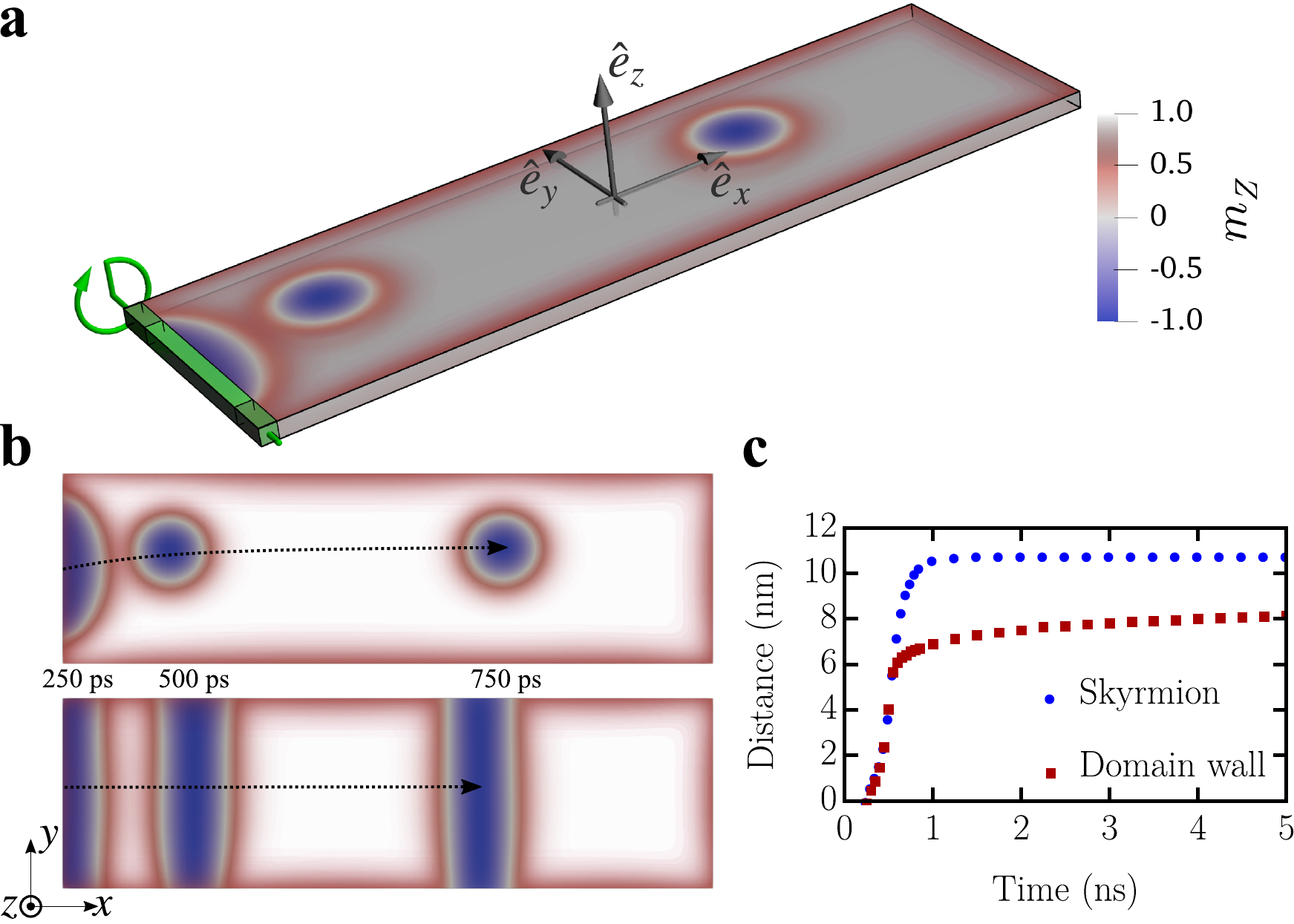}
	\caption{Skyrmion and domain wall creation in a magnetic stripe.
		(a) 
		Schematics of the rotating effective magnetic field applied locally at the edge (green rectangular area) of the system. The time-dependent magnetic field rotates along the green arrow in the $xz$-plane with angle $\theta$.
		(b)
		Real-space image of the motion of the skyrmion and DW for different time steps. 
		For the skyrmion, the field is applied to two thirds of the width of the stripe (bright green area in (a)), in the case of the DW, the full width is affected by the field. 
		The shown positions correspond to $250$~ps (half a rotation), $500$~ps (full rotation), and $750$~ps (half a period after turning off the time-dependent field). The Gilbert damping parameter is reduced to $\alpha=0.01$ in order to emphasize the differences in the motions.
		(c)
		Distance-time-diagram of a single created DW and skyrmion for a standard damping parameter $\alpha = 0.1$. 
		System size: $100\times 30\times 1 c^3$ with material parameters corresponding to  Pd/Fe/Ir(111); $c=0.233~$nm. 
	}\label{fig:F1}
\end{figure}

First, we investigate whether local magnetic fields, instead of global effects\cite{stier2017skyrmion}, can be used to create non-collinear topological magnetic quasiparticles as DWs or skyrmions. 
Several approaches could either directly or effectively realize the necessary control about the edge-fields. 
	(1) In toggle magnetoresistive random access memory (MRAM), a specific timing sequence of currents in independent signal lines smoothly rotates the writing magnetic field in micrometer-sized samples\cite{toggle2005MRAM}.  
	(2) The utilization of nanostructured magnets for obtaining smoothly rotating magnetic fields with magnitudes on the order of hundreds of mT is demonstrated in ref.\onlinecite{mcneil2010localized}.
	(3) Another possibility to generate strong local effective fields exists in tailoring multilayer systems \cite{lo2020tuning}. Lo Conte et al. stabilize room temperature skyrmions by tuning the interlayer exchange coupling between two magnetic layers. As one of the adjusted parameters is the thickness of a spacer layer between the magnetic layers, this principle could be extended to nanostructuring the spacer. Hence the strength of the interaction could be manipulated spatially.
	(4) A different procedure uses current-carrying electronic states in atoms\cite{zhang2019ultrafast} or molecules\cite{watzel2016optical}. These states host fully controllable magnetic fields when irradiated by laser fields. In the case of the atomic system, magnetic flux densities up to 47\,T are predicted\cite{zhang2019ultrafast}.
	For both the laser-excited molecular systems and the nanostructured magnetic materials, the effective magnetic fields are strongly localized and hence provide promising approaches for the local creation of nanometer-sized magnetic textures.
We simulate similar, smoothly rotating effective magnetic fields locally applied to the edge ($x=0$) of a thin magnetic stripe in the $xy$-plane and with dimensions of $100\times 30\times 1 c^3$ with $c=0.233~$nm,  as shown in Fig.~\ref{fig:F1}(a), and calculate the spin dynamics in the stripe by micromagnetic simulations and atomistic Landau-Lifshitz-Gilbert (LLG) approaches. (Details can be found in the Supplementary Material, Sec.~A \textit{[URL will be inserted by publisher])}. For the micromagnetic calculations, we conduct simulations using the open-source, GPU-accelerated software package mumax3\cite{vansteenkiste2014design},
see Methods for details.

The simulations are performed primarily with parameters for the Pd/Fe bilayer on an Ir(111) surface, known for hosting nanometer-sized skyrmions at moderate magnetic bias fields
(saturation magnetization $M\ind{sat}=1.1~$MA\,m$^{-1}$, interfacial DMI constant $D=3.9~$mJ\,m$^{-2}$, exchange stiffness $A\ind{exch}=2~$pJ\,m$^{-1}$, uniaxial anisotropy constant $K\ind{u}=2.5~$MJ\,m$^{-3}$ and Gilbert damping parameter $\alpha = 0.1$)\cite{romming2015field}. The second model system we consider is a Co/Pt multilayer
 with weaker DMI.
The local edge-field rotates in the $xz$-plane (cf. Fig~1(a)).
Additionally, a static background field of $B\ind{z}^\mathrm{stat}=1.5~$T was applied to ensure a spin-polarized ground state and the stability of both DWs and skyrmions in our calculations. 
We apply a local edge-field with an amplitude of $|\vec B (t)| = B = 10$~T, rotating once within $500~$ps ($\nu=2~$GHz) and then being deactivated, meaning 
\begin{align}
	\begin{split}
		\vec{B}(t<\nu^{-1}) &= B\sin(2\pi\nu t) \hat{\vec{e}}\ind{x} + \left[B\ind{z}^\mathrm{stat} + B\cos(2\pi\nu t)\right]\hat{\vec{e}}\ind{z}, \\
		\vec{B}(t\geq \nu^{-1}) &= B\ind{z}^\mathrm{stat} \hat{\vec{e}}\ind{z}~.
	\end{split}
\end{align}
\label{eq::Bt}
Figs.~1(b) and (c) show how stable non-collinear magnetic objects are successfully inscribed for matching rotational senses of both the field and the intrinsic magnetic chirality determined by the DMI. Chiral DWs and skyrmions can be induced by a field applied either to the complete edge (DW) or only to its central two thirds (skyrmion), respectively (Fig.~\ref{fig:F1}(b,c), cf. lighter and darker green area in Fig.~\ref{fig:F1}(a)). 
Once generated, the spin arrangements move along the racetrack until they slow down and come to rest at a finite distance from the point of creation. In order to emphasize the different dynamics of the DW and the skyrmion, the Gilbert damping parameter in Fig.~\ref{fig:F1}(b) is chosen to be relatively low ($\alpha=0.01$), to achieve significant lateral displacements. 
The positions of both objects are shown after half a rotation of the external edge-field ($250$~ps), after the full rotation ($500$~ps), and another half period after turning off the rotating effective field ($750$~ps).  The displacements of the created quasiparticles for  $\alpha=0.1$ are shown in Fig.~\ref{fig:F1}(c), an animated version can be found in the Supplementary Video~1 \textit{[URL will be inserted by publisher]}.  The path-time diagrams demonstrate that the magnetic quasiparticles possess initial momentum and inertia as they continue to move after the field has been switched off. 
As in ref.\onlinecite{schutte2014inertia}, the Thiele equation, customarily used to describe the center of mass dynamics of magnetic quasiparticles, may be extended to include higher-order terms like the gyrodamping. This is essential for our creation mechanism, as it involves highly non-linear dynamics. However, the nature of a quasiparticle model lacks the ability to treat the creation of the skyrmion itself. Therefore we employ numerical spin dynamics simulations. 

In the following, the control of the magnetic quasiparticles is performed by magnetic fields, localized at an optimized number of simulation cells close to one edge of the sample represented by a rectangular slab. This control includes displacement operations, which usually are achieved by applying global electrical currents. 
 
	In the first step, we optimized the area where the magnetic field is applied in terms of generating skyrmions or domain walls at low field amplitudes. A detailed description of this parameter optimization and the application of more sophisticated field distributions are discussed in the Methods and the Supplementary Material, Sec.~B \textit{[URL will be inserted by publisher]}.

\begin{figure}
	\centering
	\includegraphics[width=\linewidth]{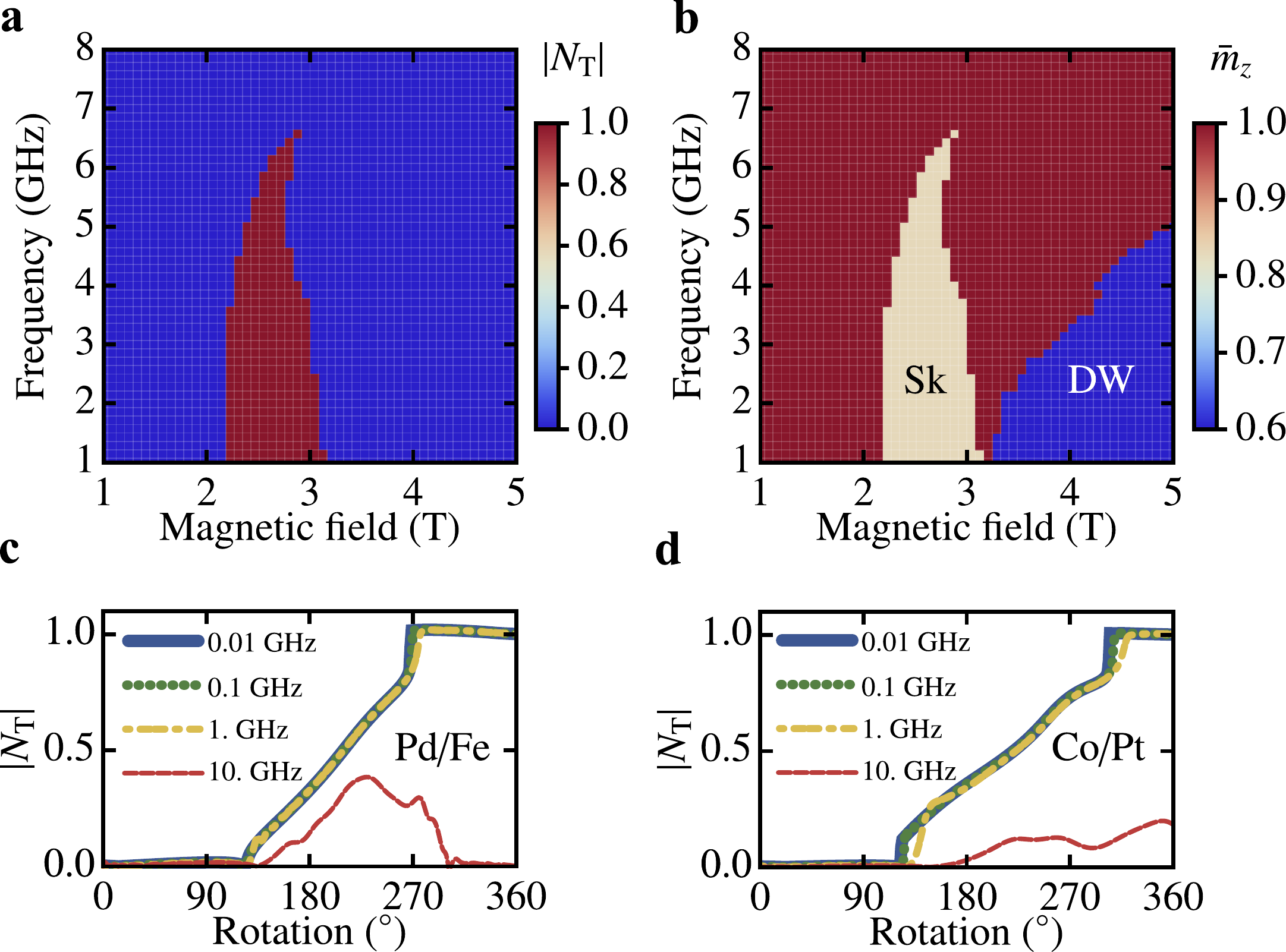}
	\caption{Quasiparticle creation due to a single rotation of a magnetic field localized at the sample edge.
		(a-b)
		Parameter space diagrams for a single field rotation and subsequent system relaxation. Shown is the absolute value of the topological charge $|N\ind{T}|$ (a), and the vertical magnetization averaged over the sample (b) as a function of the frequency and amplitude of the time-dependent driving field. The excited area at the left boundary is  $7\,c\times24\,c$ , and the damping parameter $\alpha = 0.1$. 
		(c-d) 
		Absolute value of the topological charge as a function of the rotational phase of the local field for Pd/Fe/Ir(111)(c) and Co/Pt (d). 
		System size: $60\times 30\times 1 c^3$ for (a-c) and $100\times 64\times 1 c\ind{Co}^3$ for (d), with material parameters corresponding to  Pd/Fe/Ir(111) (a-c) and Co/Pt (d); $c=0.233~$nm, $c\ind{Co}=1~$nm.  
	}
	\label{fig:F2}
\end{figure}

For a fixed area affected by the rotating effective edge-field of $7\times24$  simulation cells, the rotation speed (frequency $\nu$ from  1 to 8~GHz) and the amplitude ($B$ from  1 to 5~T) are varied. 
Based on the magnetic configurations, resulting from the rotating effective excitation field and subsequent relaxation, the absolute value of the topological charge $|N\ind{T}|$ (Fig.~\ref{fig:F2}(a)) and the out-of-plane component of the magnetization $\bar{m}_z$ averaged over the sample surface (Fig.~\ref{fig:F2}(b)) are calculated.
The topological charge $N\ind{T}$ is defined as 
\begin{align}
	N\ind{T} &= \frac{1}{4\pi}\displaystyle\int n \ind{T}(x,y)\, \dd x \dd y~, \label{eq::Nt}\\
	\intertext{where} 
	n\ind{T}&=\vec{m}\cdot\left(\frac{\partial \vec{m}}{\partial x}\times\frac{\partial\vec{m}}{\partial y}\right)
\end{align} 
is the topological charge density and $\vec{m}$ represents the vector field of the magnetization. 
As expected\cite{bottcher2018b}, we only observe skyrmions and no antiskyrmions\cite{stier2017skyrmion} in the considered parameter regime. Therefore, the sign of the topological charge only reflects the polarity of the skyrmions, which is fixed by the non-rotating edge magnetization of the sample. To avoid confusion, absolute values $|N\ind{T}|$ are shown in Fig.~\ref{fig:F2}(a). 
By combining both maps, two distinct regimes become apparent. For frequencies up to $\nu\approx 6.5~$GHz  a pocket forms in a regime of moderate field amplitudes in which a well-defined topological charge of $|N\ind{T}|=1$ is measured. This area corresponds to the injection of a single skyrmion. For larger amplitudes or higher frequencies, respectively, no skyrmion can be inscribed into the system.
Remarkably, for even higher amplitudes of the field, another distinguished area in the parameter space opens up, where the average magnetization is diminished even more than in the case of the skyrmion. It turns out that this blue area in Fig.~\ref{fig:F2}(b) represents a single created DW without  indication in the total topological charge. 
Here, the possibility of the individual creation of magnetic objects with distinct topologies opens new technological perspectives. 
Most interestingly, radially symmetric skyrmions or chiral DWs can be inscribed in the same nanostripe setup by solely changing the amplitude of the local effective rotating edge-field.  This is in contrast to Figs.~\ref{fig:F1}(b) and (c), where we assumed different spatial extensions of the magnetic fields. 

As a second model system we use Co/Pt multilayers, in which skyrmions have been observed in the absence of global magnetic fields ($M\ind{sat}=0.58~\mathrm{MA}\,\mathrm{m}^{-1}$, $A\ind{exch}=15~\mathrm{pJ}\,\mathrm{m}^{-1}$, $D=3.5~\mathrm{mJ}\,\mathrm{m}^{-2}$, $K\ind{u}=0.8~\mathrm{MJ}\,\mathrm{m}^{-3}$, $\alpha=0.3$, system size: $100\times 64\times 1~$nm$^3$, cell size: $1~$nm$^3$)\cite{zhang2016control,sampaio2013nucleation}. In Fig.~\ref{fig:F2}(c) and (d) the topological charge is displayed for both materials as a function of the rotation angle $\theta$ (cf. Fig.~\ref{fig:F1}(a)) of the external field. The rotation frequencies span from $10$~MHz up to $10$~GHz for an amplitude of the localized rotating  effective  field of  $B=2.5$~T in the case of the Pd/Fe/Ir(111) system. For Co/Pt an area of $7\times 48$~simulation cells is excited with a field amplitude of $1~$T (see Methods for details on the choice of parameters).  
Note that due to the non uniform magnetization of the boundary, the topological charge varies smoothly.
New quasiparticles are created at rotation angles  $\Delta \varphi_\mathrm{Pd/Fe} \approx 270^\circ$; $\Delta \varphi_\mathrm{Co/Pt} \approx 300^\circ$  (where $|N_T|$ reaches unity in Fig.~\ref{fig:F2}(c-d)).
Though the material parameters differ significantly and we do not apply a static magnetic field for stabilizing skyrmions in the case of Co/Pt, the quasiparticle creation is still successful in a similar frequency regime up to the GHz range (Fig.~\ref{fig:F2}(d)). 
This indicates that our results are not limited to the Pd/Fe/Ir(111) system, but can be applied to various chiral magnets. 

\begin{figure}
	\centering
	\includegraphics[width=\linewidth]{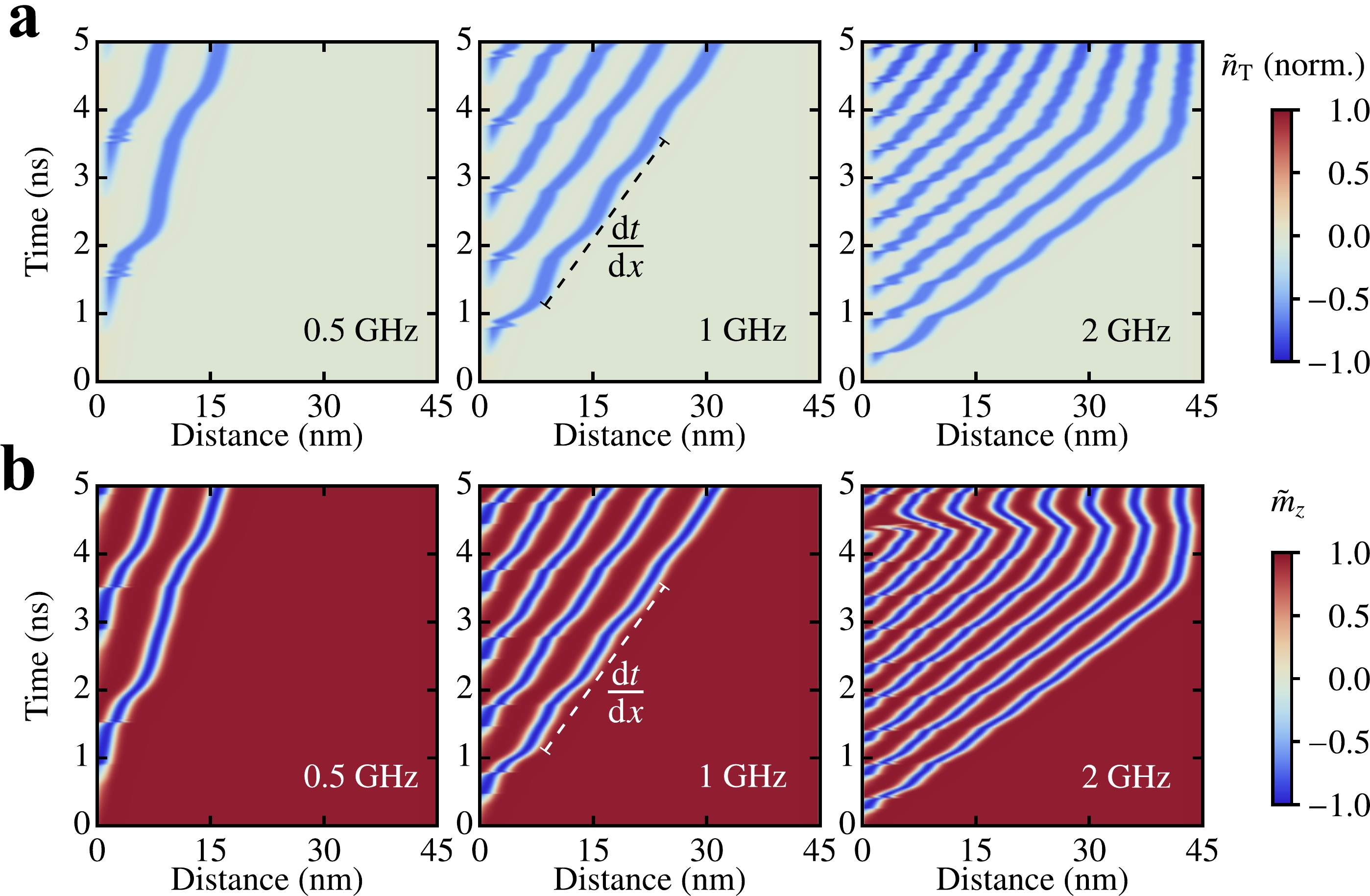}
	\caption{Skyrmion and domain wall steered by a continuously rotating effective magnetic field localized at the sample edge.
		Path-time diagrams of the skyrmion (a) and DW (b) propagation for $\nu=0.5,1,2$ GHz, respectively (from left to right). Each horizontal line of pixels shows (a) the topological charge density or (b) the out-of-plane component of the magnetization, each averaged over cross sections of the nanostripe in ($\tilde{n}\ind{T}(x,t)$, $\tilde{m}\ind{z}(x,t)$). 
		The inverse mean slopes of the curves give the velocities of the quasiparticles.  
		System size: $200\times 30\times 1 c^3$, with material parameters corresponding to  Pd/Fe/Ir(111); $c=0.233~$nm.}
	\label{fig:F3}
\end{figure}

In the next step, we investigate whether a created magnetic object can be moved to any position along the racetrack using local effective rotating edge-fields. To achieve this goal, continuously rotating fields of different frequencies are applied to the Pd/Fe/Ir(111) nanostripe edge. Here, the excited area is again  $7\times 24$ simulation cells large , whereas the amplitude of the field is set to $B\ind{Sk}=2.5~$T or $B\ind{DW}=3.5~$T according to Fig.~\ref{fig:F2}(b). The path-time diagrams are shown in Fig.~\ref{fig:F3} for both the topological charge density in the case of the skyrmion creation (Fig.~\ref{fig:F3}(a)) and the out-of-plane component of the magnetization for DW simulations (Fig.~\ref{fig:F3}(b)). Both quantities are averaged over one cross-section of the stripe  $\tilde{n}\ind{T}(x)=\langle n\ind{T}(x,y) \rangle_y$~, $\tilde{m}_z(x)=\langle m_z(x,y) \rangle_y$~ to track the skyrmions and DWs along the stripe more easily. 
For sufficiently large DMI, the created magnetic objects are stable and obtain an initial velocity that drives them away from the boundary (see Fig.~\ref{fig:F3}(a-b)). In contrast to racetrack concepts based on electrical currents\cite{parkin2008magnetic,tomasello2014strategy,gobel2019electrical}, where a single magnetic object is moved along a racetrack, in our setup, trains of quasiparticles are created and moved collectively: at each subsequent $2\pi$ rotation of a magnetic field, a new quasiparticle is created, while already existing quasiparticles are moved away due to the repulsive interactions between them (see Fig.~\ref{fig:F3}(a-b)). 
Furthermore, the distance traveled by the created quasiparticles during the creation time increases linearly with increasing frequency. This can be seen in the path-time diagrams of Fig.~\ref{fig:F3}(a-b), where the inverse of the linear slope gives the mean velocity of the quasiparticle, which reaches $v\approx 12~$m\,s$^{-1}$ at $\nu = 2~$GHz. This speed can be further increased by smart combinations of material parameters and frequencies of the rotating effective field. 
The distances traveled by each of the particles suggests a linear speed-frequency relation. 
Additionally, the minor delay in the motion of previously and newly created quasiparticles in the stripe implies that the time scale of magnetic interactions is much smaller than that of the driving local edge-field. 
This limits the validity of the suggested linear speed-frequency relation to sufficiently low frequencies or large damping parameters.  
The velocity of propagation decreases when the quasiparticles fill the sample (see right panels of Fig.~\ref{fig:F3}(a-b) for $\nu=2~$GHz). In this regime, the mean distance of skyrmions/DWs is comparable to the interaction range of these objects\cite{schaffer2019stochastic,pinna2018skyrmion,lin2013particle,reichhardt2018clogging}. 
If the energy barrier given by the repulsion of the quasiparticle from the edge is large enough, the packing density of quasiparticles increases as shown in detail in the Supplementary Video~2 \textit{[URL will be inserted by publisher]}.  
 A difference between skyrmion and DW creation starts to appear at high quasiparticle densities, i.e., in the 2~GHz panel for $t>4~$ns. Here the skyrmion injection persists, whereas a maximal possible density of DWs is reached and excess DWs are repelled out of the sample after a short backaction. This behavior can be related to the smaller feature size of a skyrmion compared to a DW.

%
%

\begin{figure}
	\centering
	\includegraphics[width=\linewidth]{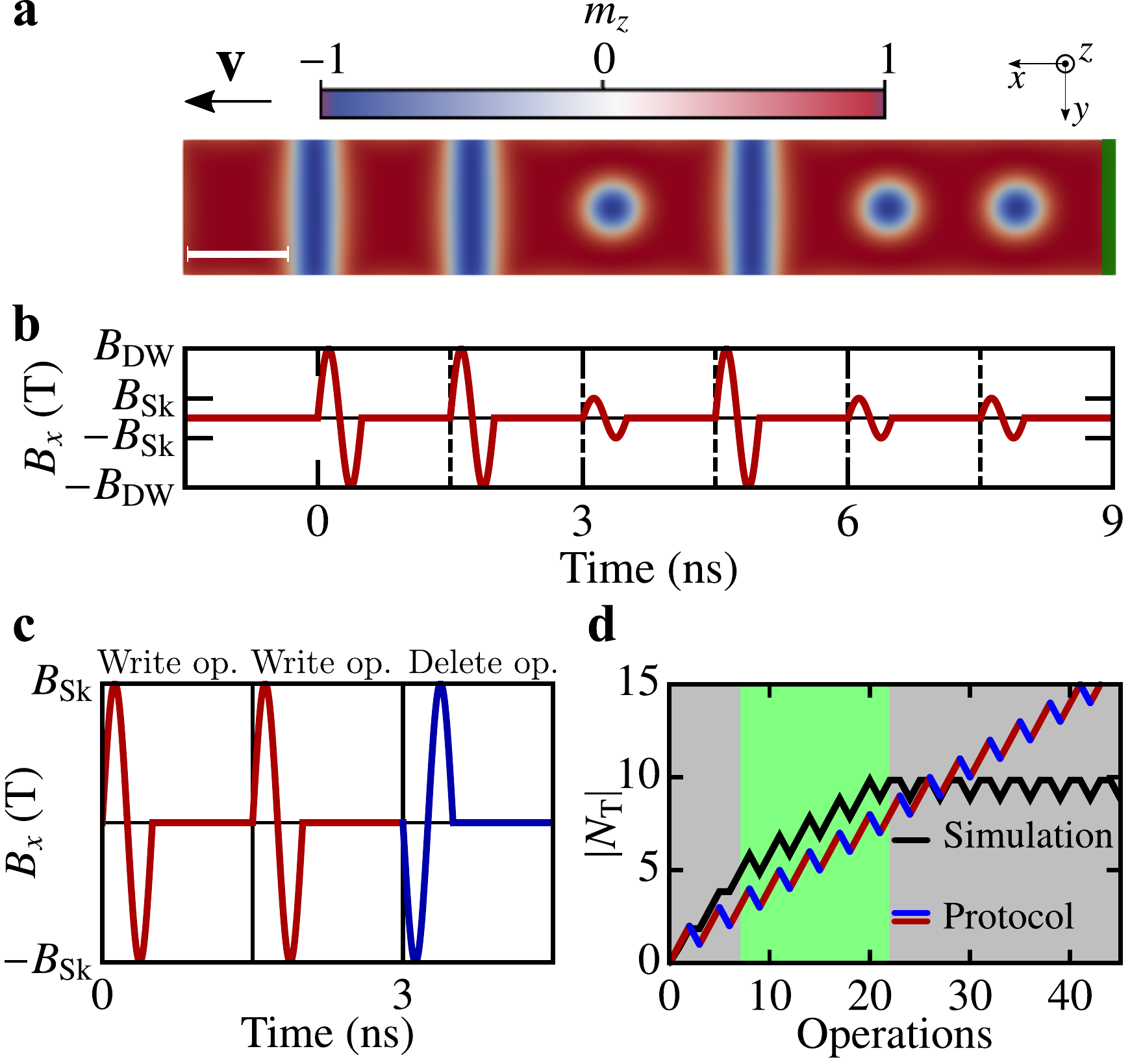}
	\caption{Writing and deleting operations.
		(a)	
		Mixed sequence of three DWs and three skyrmions generated by multiple different field operations in the green area. 
		(b)	
		In-plane component of the rotating effective magnetic field with two different amplitudes. 
		(c)	
		Field protocol $B\ind{x}(t)$ for two writing operations (red) and one deletion operation (blue) with rotational sense opposite to the DMI. 
		(d) 
		Simulated total topological charge (black line) as a function of the number of subsequent field operations according to the protocol shown in (c). In the bright green area, the simulated topological charge is parallel to the behavior of an ideal system exactly following the protocol (red/blue line), therefore indicating high reproducibility.
		System size: $200\times 30\times 1 c^3$, with material parameters corresponding to  Pd/Fe/Ir(111).}
	\label{fig:F4}
\end{figure}

So far, we have shown that it is possible to create and move trains of magnetic quasiparticles by local excitations in a controllable way, without currents or global driving fields. Only the local edge-field is necessary for creating an object, the intrinsic interactions result in quasiparticle propagation. Because the DMI has a directional sense and because the quasiparticles repel each other, the creation of an additional object lets the existing quasiparticles move in a given direction with a well-defined velocity.
To increase the functionality of the proposed current-free racetrack, we check the possibility of using magnetic quasiparticles of different topology as bits of information based on our findings in Figs.~\ref{fig:F2}(a-b). Fig.~\ref{fig:F4} shows results for different combinations of effective  rotating edge-fields.
Fig.~\ref{fig:F4}(a) is obtained by changing the amplitude of the driving edge-field according to the protocol shown in Fig.~\ref{fig:F4}(b) for the $x$ component of the magnetization (cf. equation~\ref{eq::Bt}(1)). 
Each operation, separated by a dashed line in the graphic representation, consists of a single rotation of amplitude $B\ind{Sk}$ or $B\ind{DW}$ and two periods (in units of $\nu^{-1}$) of relaxation time with $\nu =2~$GHz (see Supplementary Video~3\textit{[URL will be inserted by publisher]} for an animated version of the resulting magnetization dynamics). Thus, the creation and successful propagation of any mixed sequence of DWs and skyrmions are possible. 

Finally, we want to study the dynamics of the created magnetic objects if the rotational sense of the local field is reversed to potentially delete previously created quasiparticles. 
Fig.~\ref{fig:F4}(c) shows the field operations considered for studying the necessary skyrmion densities to permit reversible creation and annihilation operations. Starting from a field polarized state, two writing attempts and one deletion process with opposite chirality are performed (see Supplementary Video~4\textit{[URL will be inserted by publisher]} for an animated version of the resulting magnetization dynamics in a small model system). 
The resulting topological charge of the Pd/Fe stripe after each operation is shown in Fig.~\ref{fig:F4}(d) along with the ideal response according to the discussed protocol.
Above a threshold topological charge of $|N\ind{T}|\gtrsim   10$, the skyrmion density is too large to add further quasiparticles successfully. Instead, undesired escapes of skyrmions take place (second gray area in Fig.~\ref{fig:F4}(d)).
In contrast, for low skyrmion densities ($|N\ind{T}|\lesssim 5$) the quasiparticles lose contact to the boundary and can not be deleted reliably (first gray area in Fig.~\ref{fig:F4}(d)). 
In the bright green region, the field protocol is parallel to the writing-deleting events; hence, this area signifies high operational stability. 

\section*{Discussion}
Global current-free writing, propulsion, and deletion of magnetic quasiparticles with distinct topological properties can be achieved in the same system by local excitations at the boundary of a magnetic racetrack device. The proposed procedure relies on internal magnetic interactions instead of global driving currents, which strongly decreases the required energy consumption and potentially avoids overheating.
By attributing a binary zero to one of the two topologically distinct objects, while binary unity to the other, a racetrack memory can be developed in a suitable quasiparticle density regime. 
The main benefit of utilizing two different kinds of quasiparticles is avoiding relying on a conserved void between objects of a single species. Due to the thermal motion of magnetic quasiparticles\cite{schaffer2019stochastic}, they naturally distribute uniformly over samples at finite temperatures. Previously, different approaches have been developed to evade this problem, e.g., by periodically modifying the energy surface\cite{suess2018repulsive} such that objects rest at preferential positions. Therefore, the quasiparticles can no longer move isotropically, and, once written, information encoded in occupied and unoccupied positions is conserved. Therefore increased energy is needed to move objects at all and the necessary precise tuning of driving forces to move quasiparticles to the desired positions. 
These challenges do not apply to the concept presented here, as the placeholder between magnetic objects as bits of information is another magnetic quasiparticle. 
Additionally, the current-free storage concept proposed here is compatible with a three-dimensional design of magnetic networks, which significantly increases the information storage capacity.
\section*{Methods}
\subparagraph{Micromagnetic simulations}
	The functional derivative of the free energy density $\mathcal{F}[\vec{m}]$ with respect to the unit vector field of the magnetization $\vec{m}(\vec{r},t)$ defines the time- and space-dependent effective magnetic field
	\begin{equation}
	\vec{B}_i^{\mathrm{eff}}(t)=\vec{B}_i^\mathrm{ext}(t)+\vec{B}_i^\mathrm{exch}+\vec{B}_i^\mathrm{d}+\vec{B}_i^\mathrm{a}+\vec{B}_i^\mathrm{dmi}~. \label{eq:heff}
	\end{equation}
	It is composed of the external field $\vec{B}_i^\mathrm{ext}(t)$, including the rotating edge-field; the exchange interaction field  $\vec{B}_i^\mathrm{exch}=2A\ind{exch}/M\ind{sat}\Delta\vec{m}_i$, with the exchange stiffness $A\ind{exch}$ and the saturation magnetization $M\ind{sat}$, the demagnetizing field $\vec{B}_i^\mathrm{d}=M\ind{sat}\hat{\vec{K}}_{ij}*\vec{m}_j$, where we refer to ref.\onlinecite{vansteenkiste2014design} for details of the calculation of the demagnetizing kernel $\hat{\vec{K}}$, the uniaxial magnetocrystalline anisotropy field $\vec{B}_i^\mathrm{a}=2K\ind{u}/M\ind{sat} m_z\vec{e}_z$, with $K\ind{u}$ the anisotropy constant, and the field generated by the Dzyaloshinskii-Moriya interaction $\vec{B}_i^\mathrm{dmi}=2D/M\ind
	{sat}\left(\partial_x m_z,\partial_y m_z,-\partial_x m_x - \partial_y m_y\right)^T$, with $D$ the strength of the interfacial Dzyaloshinskii-Moriya interaction. 
	The effective magnetic field enters the LLG equation,
	\begin{equation}
	\dot{\vec{m}}_i(t)=-\frac{\gamma}{1+\alpha^2}\left[\vec{m}_i\times\vec{B}_{i}^\mathrm{eff}+\alpha\vec{m}_i\times\left(\vec{m}_i\times\vec{B}_{i}^\mathrm{eff}\right) \right]~,
	\end{equation}
	which is solved for every simulation cell $i$ of the discretized magnetization vector field $\vec{m}_i$ . 
	The gyromagnetic ratio of an electron is denoted by $\gamma_0=1.76\times 10^{11}($T$^{-1}$s$^{-1})$ and $\alpha$ is the Gilbert damping parameter. 
	The integral definition of the topological charge $N\ind{T}$ (Eq.\ref{eq::Nt}) is applied to the discrete lattice system.

\subparagraph{Optimization of excitation area}
	To optimize the excitation mechanism for the uniformly distributed magnetic field restricted to a rectangular shaped area, two benchmarks to rate magnetic field sweeps are defined. 
	First, the possibility of choosing between skyrmion or DW creation by changing the amplitude of the rotating magnetic field should be preserved. Hence we are looking for separate regimes of skyrmion and DW creation. 
	Second, the proposed device should work at minimum field amplitudes to ensure low energy consumption.  \\
	For Pd/Fe/Ir(111) the excited area is varied between $(1\times 2)$ and $(15\times 30)$ simulation cells for a fixed frequency of $\nu=1~$GHz. From the obtained 240 B-field sweeps, 53 show both features of skyrmions (topological charge $|N\ind{T}|=1$) and DWs (average out-of-plane magnetization $\bar{m}_z<0.7$). Afterward, the data set with the lowest threshold magnetic field for the DW creation is selected from the filtered data sets. This value is favored over the critical skyrmion creation field, as it is the highest necessary field for operating the mixed skyrmion DW track. The scan through frequency and field amplitude for the chosen area of excitation of $(7\times 24)$ simulation cells is shown in Fig.~2(a,b) of the main text. \\
	The same procedure is applied for the Co/Pt system, where the excited area is altered between $(1\times 2)$ and $(52\times 64)$ simulation cells. Here 32 out of the 352 data sets show indications for both DWs and skyrmions, and we identified an area of $(7\times 48)$~simulation cells for optimal functionality. 
	
	%

%

\section*{Acknowledgements}
Financial support provided by the Deutsche Forschungsgemeinschaft (DFG) via the Cluster of Excellence "Advanced Imaging of Matter" (DFG EXC 2056, project no. 390715994), CRC/TRR 227 and within the SPP 2137 "Skyrmionics" (project no. 403505707) is gratefully acknowledged.

\section*{Author Contributions}
E.Y.V. proposed and developed a general concept of this investigation. A.F.S. conducted and analyzed the micromagnetic simulations. P.S., M.S. and E.Y.V. performed the atomistic spin dynamics simulations. A.F.S. and E.Y.V. wrote the manuscript. 
R.W., M.T., J.B., T.P., M.S. and P.S. discussed the results and contributed to the manuscript. 
\\

\end{document}


\title{Supplementary material: Rotating edge-field driven processing of chiral spin textures in racetrack devices}

	
	\author{Alexander F. Sch\"affer}
	\affiliation{Institute of Physics, Martin-Luther-Universit\"at Halle-Wittenberg, D-06120 Halle (Saale), Germany}
	\affiliation{Department of Physics, Universit\"at Hamburg, D-20355 Hamburg, Germany}
	
	\author{Pia Siegl}
	\author{Martin Stier}
	\author{Thore Posske}
	\affiliation{I. Institute for Theoretical Physics, Universit\"at Hamburg, D-20355 Hamburg, Germany}
	
	\author{Jamal Berakdar}
	\affiliation{Institute of Physics, Martin-Luther-Universit\"at Halle-Wittenberg, D-06120 Halle (Saale), Germany}
	
	\author{Michael Thorwart}
	\affiliation{I. Institute for Theoretical Physics, Universit\"at Hamburg, D-20355 Hamburg, Germany}
	
	\author{Roland Wiesendanger}
	\author{Elena Y. Vedmedenko}
	\affiliation{Department of Physics, Universit\"at Hamburg, D-20355 Hamburg, Germany}

	\maketitle

%
%
%
%
%
%
%
%
%

\section{Atomistic simulations on a monolayer with an hcp(111) stacking}
	For the systems, discussed in the main text, we have used a discretization into cubic simulation cells. However, ultrathin films such as the Pd/Fe bilayer on Ir(111) typically grow in hexagonal close-packed (hcp) stacking.
	Here, we strictly use the hcp structure of magnetic adatoms and show the applicability of the presented writing and deleting mechanism also to these cases.
	Additionally, the skyrmions are stabilized by a combination of the DMI and the geometric confinement entering as fixed boundary conditions. This represents another example of nanometer sized skyrmions in the absence of global magnetic fields.
	
	For the atomistic simulations, classical Heisenberg magnetic moments
	$\vec{S_i}=({S_i}^x,{S_i}^y,{S_i}^z)$ of unit length $\vec\mu_i/\mu\ind{s}$ are placed at each lattice point.
	The energy of the system is given by the Hamiltonian
	\begin{equation}
	\label{e:hamilton}
		\mathcal{H} = -J \sum_{\langle ij\rangle} \vec{S}_{i}\cdot \vec{S}_{j}
		- \sum_{\langle ij\rangle} \vec D_{ij}\cdot (\vec S_{i}\times \vec S_{j})-B_z\sum_{i}S_z-\sum_{i}\vec B_i(t)\cdot \vec S_i- K_z\sum_i (S_{i}^z)^2,
	\end{equation}
	where $J>0$ is the ferromagnetic exchange coupling between nearest neighbors, $\vec D_{ij}$ is the DMI vector, $B_z$ is a global magnetic field, $K_z$ is the perpendicular magnetic anisotropy and $\vec B_i(t)$ is a local space- and time-dependent magnetic field.
	The material parameters were set to $K_z=0.05\,J$, $D=0.07\,J$ and the magnetic moments at the edges are fixed to $\vec{S}=\hat{\vec{e}}_z$ to stabilize skyrmions in the absence of an additional external field. The spins at $x=0$ were rotated parametrically. The time scale in these calculations can be expressed by $dt=\tau \mu_s/\gamma J$ with the reduced time step $\tau$ and the gyromagnetic ratio $\gamma$. For $\mu_s=2\mu_B$, $J=5.72$ meV and $\tau=0.01$ a simulation step corresponds to the time span of $\approx 10^{-14}$ s. The frequency of rotation at the edge of the stripe was varied between 1 and 10 GHz. The propagation velocities were similar to those found in micromagnetic simulations.	
	\\
	The dynamics of the free spins is determined by the LLG equation
	\begin{equation}
		\dot{\vec{m}}_i(t)=-\frac{\gamma}{1+\alpha^2}\left[\vec{m}_i\times\vec{B}_{i}^\mathrm{eff}+\alpha\vec{m}_i\times\left(\vec{m}_i\times\vec{B}_{i}^\mathrm{eff}\right) \right]~,
	\end{equation}
	with the effective magnetic field
	\begin{equation}
		\vec{B}_i^\mathrm{eff} = -\frac{\partial \mathcal{H}}{\mu_s \partial \vec{S}_i}~.
	\end{equation}
	
\begin{figure}
		\includegraphics*[width=0.6 \columnwidth]{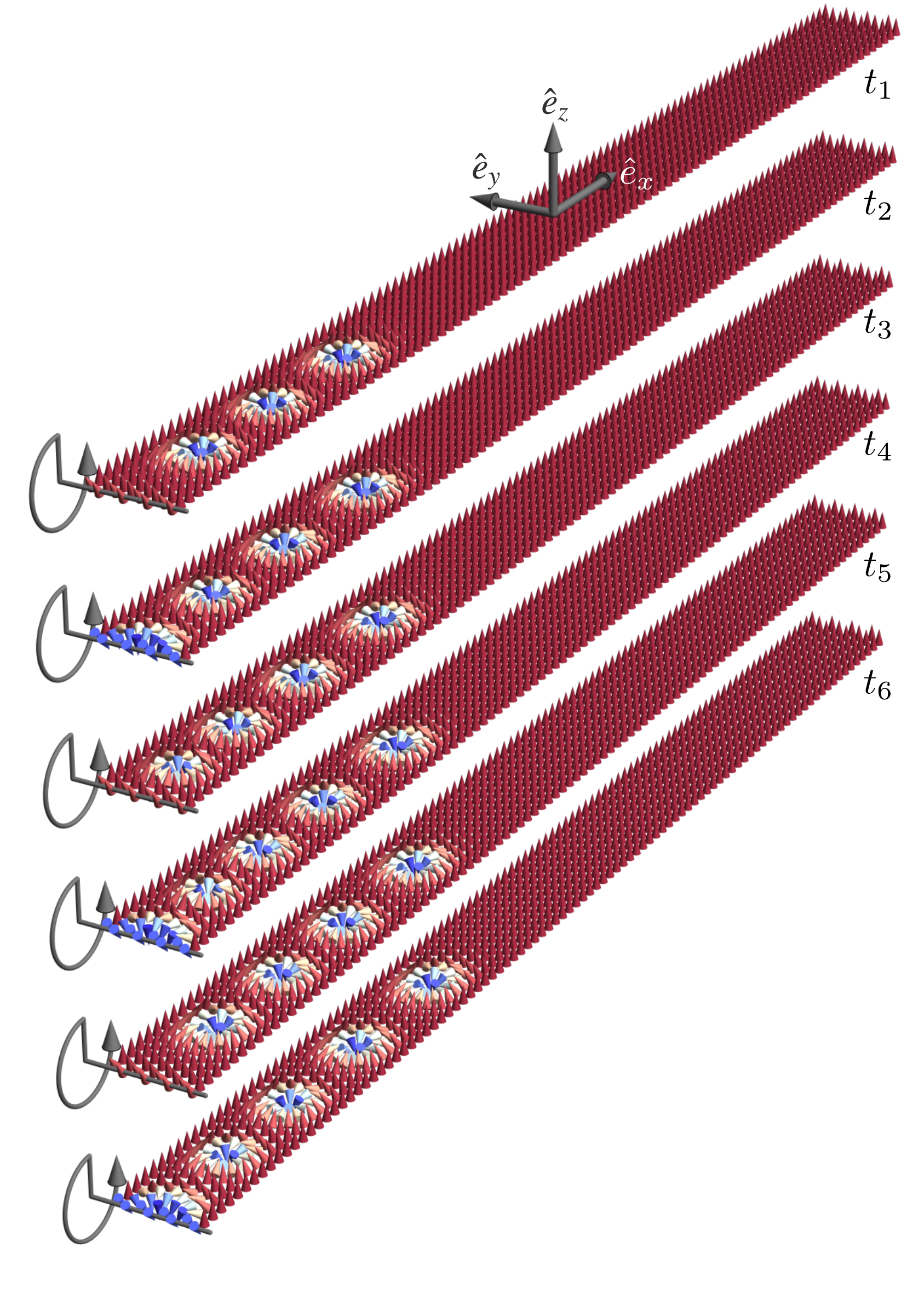} 
		\caption{
			Skyrmion creation in a magnetic stripe from atomistic simulations.
			Snapshots corresponding to the writing of individual skyrmions by a local rotating edge-field following the chirality of Dzyaloshinskii-Moriya interaction.
		}
		\label{fig:figS1a}
	\end{figure}

\begin{figure}
		\includegraphics*[width=0.6 \columnwidth]{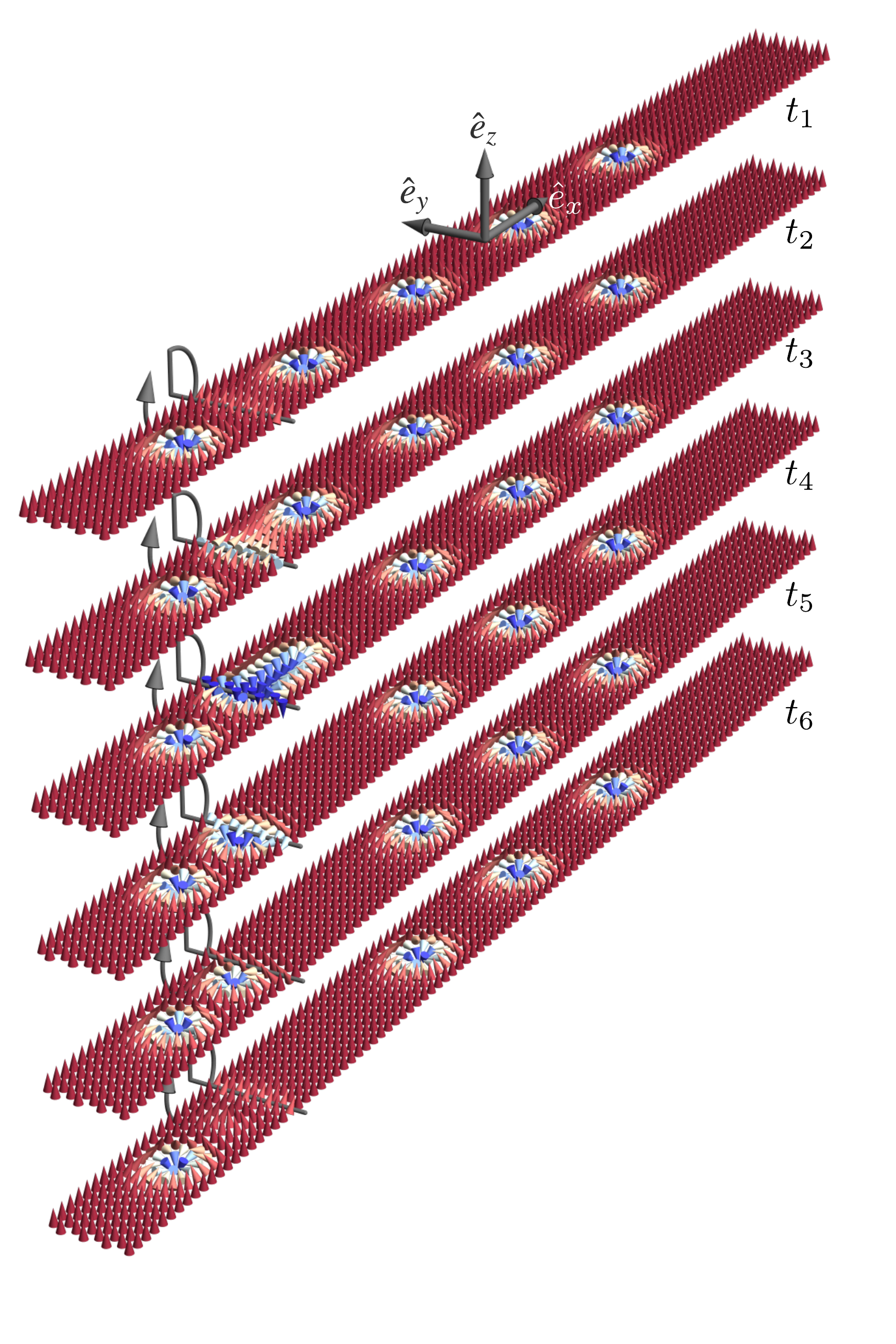} 
		\caption{
			Skyrmion deletion in a magnetic stripe from atomistic simulations.
			Local rotating edge-field opposite to the rotational sense in Fig.~S\ref{fig:figS1a} leads to the deletion of a skyrmion in $+x$ direction.
		}
		\label{fig:figS1b}
	\end{figure}

	Fig.~S\ref{fig:figS1a} shows a sequence of snapshots of the creation and deletion of magnetic skyrmions using atomistic LLG simulations on a triangular lattice with $8\times 100$~ spins.
	The excited line of magnetic moments is visualized by the gray screw, also indicating the rotational sense.
	As shown in the micromagnetic simulations presented in the main text, skyrmions can be injected into the sample.
	In Fig.~S\ref{fig:figS1b} the spins along the screw are rotated in opposite direction. Particularly interesting is that the rotation against the DMI chirality destroys skyrmions only on one side of the excited line as can be seen in time steps $t_3$ to $t_6$ of Fig.~S\ref{fig:figS1b}.
	
	In conclusion, this shows that skyrmions stabilized by DMI and fixed boundary conditions can be created and annihilated similar to the open systems discussed in the main text. Furthermore, the hexagonal lattice has no fundamental impact on the proposed control mechanism.
	

\clearpage

\section{Phase diagram of skyrmion creation for Gaussian rotation}
The creation processes of either domain walls or skyrmions discussed in the main paper have been modeled by a simplified rotation scheme, namely a uniform rotation of a distinct number of involved edge magnetic moments due to a rotating magnetic field. While for domain walls this intuitively is the optimal scheme, a rotation respecting the intrinsic circular form of skyrmions is a more adequate skyrmion creation process. As we will show in the following, such a tailored rotation allows for higher creation rates, which in the context of information technologies is a desirable goal. \\
We study the general properties of the skyrmion creation and compare different rotation schemes of a model system. To this end, we consider a square lattice of 128 x 128 lattice sites with ferromagnetic boundary conditions and system parameters of {$A\ind{exch} = 0.69~$pJ~m$^{-1}$, $D = 0.267~$mJ~m$^{-2}$ and $B = 0.0864$ T. Here, the ground state is the ferromagnetic state, with all magnetic moments pointing in $+z$-direction. Furthermore, we directly rotate the magnetic moments at the edge of the sample at $x=0$  which reflects the limits of sufficiently large magnetic fields.
	Considering the uniform rotation of all involved boundary moments, the maximum rotation frequency resulting in a stable skyrmion depends on the number of rotated edge moments, see Fig.~S\ref{rotation_scheme}(b). An optimum is found for 20 to 60 rotated moments where a rotation with maximum frequency of $\nu=12.5$ GHz can create a stable skyrmion.
	\\
	A skyrmion creation by a uniform rotation is only possible due to the directional sense of the DMI inclining the outer moments towards the outside, such that a circular structure is formed. The creation speed is therefore determined by this process. However, it can be increased if the inclination of the moments is itself induced by the rotation. A significant improvement is achieved by a tailor-made rotation which we call the Gauss rotation.
	The name and rotation protocol are motivated by the z-component of the normalized magnetization $m_z$ of the cross section of a skyrmion which can approximately be fitted by a shifted Gauss function, i.e.
	\begin{equation}
		\label{eq::B1}
		m_z(y)= -2\exp\left[-\left(\frac{y-y_0}{\sigma}\right)^2\right]+1.
	\end{equation}
	Here $\sigma$ is a measure of the width of the Gauss distribution, $y$ refers to the position of the 
	magnetic moment and $y_0$ to the center of the skyrmion, coinciding in the following with the center of the edge.
	Fig.~S\ref{rotation_scheme}(a) illustrates the scheme of the Gauss rotation.
	The moments (black arrows) rotate around the spatially dependent rotation axis (red arrow) and are depicted in their initial state (dashed line) and after half a rotation (continuous line). The angle $\varphi$ between $\textbf{m}(t=0)
	$ and $\textbf{m}(t=\frac{T}{2})$, T being the period of the rotation, depends on the Gauss function in equation~\ref{eq::B2}, namely 
	\begin{equation}
			\label{eq::B2}
			\cos(\varphi(y))= -2\exp\left[-\left(\frac{y-y_0}{\sigma}\right)^2\right]+1.
	\end{equation}
	Using this rotation, we find the creation diagram of the Gauss rotation, see Fig.~S\ref{rotation_scheme}(c).
	The Gauss rotation yields the possibility of skyrmion creation at significantly higher frequencies up to 20.8 GHz, i.e., a creation of a skyrmion within 50 ps, for an optimal Gauss width $\sigma=20$. This underlines the increased efficiency of the rotation, if it adequately forms the skyrmion shape. 
	
	\begin{figure}[!htbp]\centering
		\includegraphics[width=\linewidth, height=\textheight,keepaspectratio]{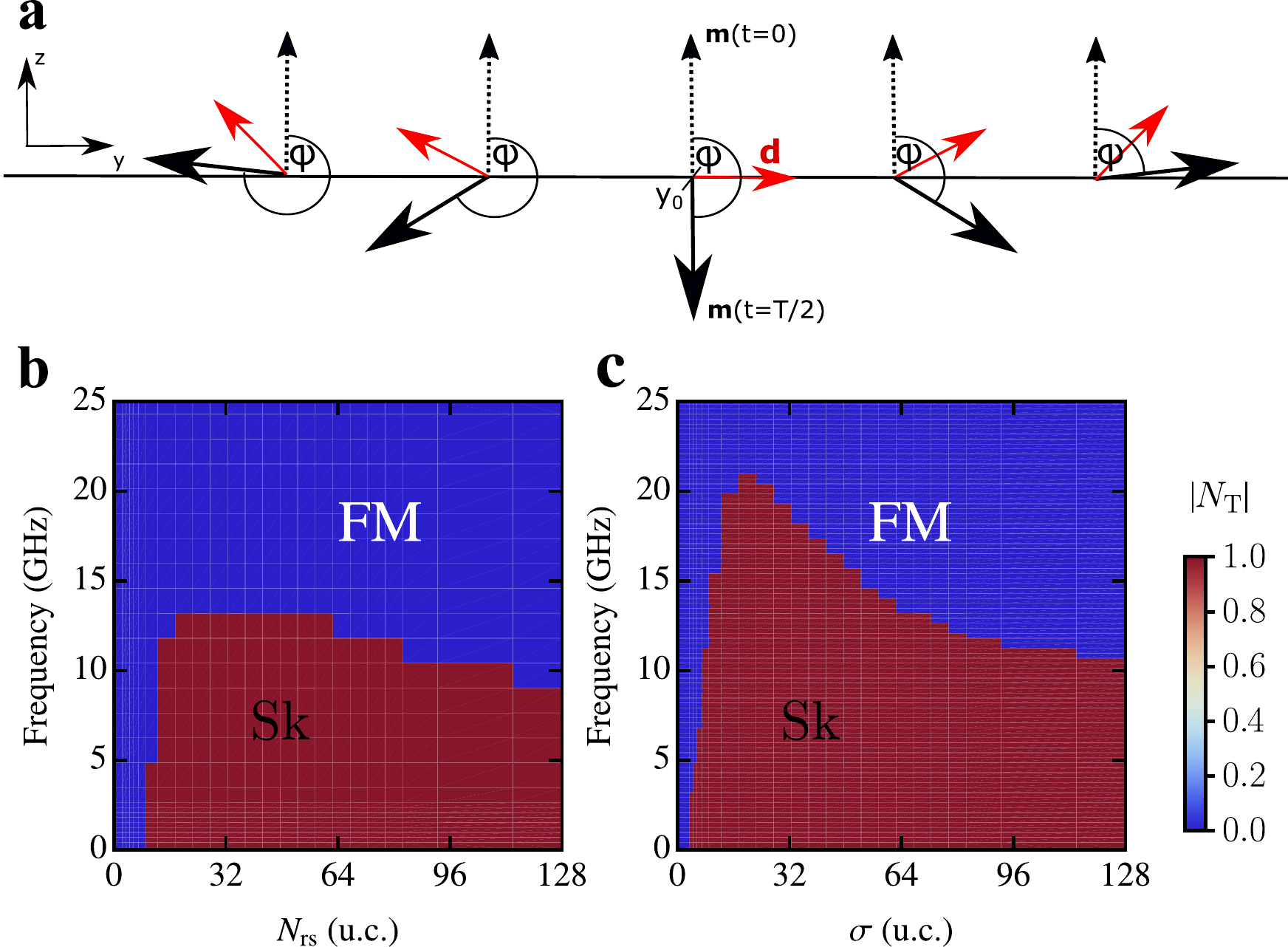}
		\caption{
			(a) Scheme of the Gauss rotation. The magnetization $\textbf{m}$ is depicted for the initial state (black dashed arrow) and after half a rotation (black arrow) around the spatially dependent rotation axis (red arrow). 
			(b,c) Phase diagrams of the creation of a single skyrmion by either the uniform (b) rotation or the Gauss rotation (c). Depicted is the absolute value of the topological charge in dependence of the number of the rotated magnetic moments $N\ind{rs}$ (uniform rotation) or the Gaussian width  $\sigma$ (Gauss rotation) and the rotation frequency $\nu$. The skyrmion is created at the end of a complete rotation, but as some will annihilate, the diagram depicts the stable solution after a sufficiently large time after the rotation.}
		\label{rotation_scheme}
	\end{figure}